# Epitaxial growth of high quality Mn$_3$Sn thin films by pulsed laser deposition

Dong Gao[1,2], Zheng Peng[1,2], Ningbin Zhang[3,4,5]*, Yunfei Xie[1,2], Yucong Yang[1,2], Weihao Yang[1,2], Shuang Xia[1,2], Wei Yan[1,2], Longjiang Deng[1], Tao Liu[1,2], Jun Qin[1,2], Xiaoyan Zhong[3,4,5], Lei Bi[1,2]*

[1] National Engineering Research Center of Electromagnetic Radiation Control Materials, University of Electronic Science and Technology of China, Chengdu 610054, China

[2] School of Electronic Science and Engineering, University of Electronic Science and Technology of China, Chengdu 610054, China

[3] TRACE EM Unit and Department of Materials Science and Engineering, City University of Hong Kong, Kowloon 999077, Hong Kong SAR, People's Republic of China.

[4] City University of Hong Kong Shenzhen Futian Research Institute, Shenzhen 518048, People's Republic of China.

[5] Nanomanufacturing Laboratory (NML), City University of Hong Kong Shenzhen Research Institute, Shenzhen 518057, People's Republic of China.

*Corresponding author.

KEYWORDS: *Mn$_3$Sn; Magnetic Weyl semimetal; Epitaxial films; Pulsed laser deposition; Magneto-optical*

ABSTRACT: Non-collinear antiferromagnet Weyl semimetal Mn₃Sn have attracted great research interest recently. Although large anomalous Hall effect, anomalous Nernst effect and magneto-optical effect have been observed in Mn₃Sn, most studies are based on single crystals. So far, it is still challenging to grow high quality epitaxial Mn₃Sn thin films with transport and optical properties comparable to their single crystal counterparts. Here, we report the structure, magneto-optical and transport properties of epitaxial Mn₃Sn thin films fabricated by pulsed laser deposition (PLD). Highly oriented $Mn_{3+x}Sn_{1-x}$ (0001) and (11$\bar{2}$0) epitaxial films are successfully growth on single crystalline Al₂O₃ and MgO substrates. Large anomalous Hall effect (AHE) up to $|\Delta R_H| = 3.02\ \mu\Omega\ \text{cm}$, and longitudinal magneto-optical Kerr effect (LMOKE) with $|\theta_K|$ = 38.1 mdeg at 633 nm wavelength are measured at 300 K temperature, which are comparable to Mn₃Sn single crystals. Our work demonstrates that high quality Mn₃Sn epitaxial thin films can be fabricated by PLD, paving the way for future device applications.

**INTRODUCTION**

Magnetic Weyl Semimetals (WSM) featuring unique band structure, enhanced anomalous Hall effect and magneto-optical effect have attracted great research interest recently. In particular, $Mn_3X$ (X=Sn, Ge, Ir) materials showing noncollinear antiferromagnetic spin texture and magnetic WSM phase at room temperature have been widely studied.[1-3] $Mn_3Sn$ is a (0001) plane ABAB stacking sequence hexagonal antiferromagnetic (AFM) material, composed of a Kagome lattice of Mn magnetic moment.[4, 5] The mixture of inter-site AFM and Dzyaloshinskii-Moriya interactions (DMI) brings about an inverse triangular spin structure below the Néel temperature $T_N$ ≈ 430 K. There are three-sublattice antiferromagnetic states on the Kagome bilayers, which are considered to be the ferroic ordering of cluster magnetic octupole.[6,7] A slightly distorted Kagome lattice constitutes each of A-B plane, and its Mn moment is ~3 $\mu_B$. The geometrical frustration leads to an inverse triangular spin structure, which shows very small net ferromagnetic moment.[8] Importantly, although their vanishingly small net magnetization, it exhibits a large AHE, which is due to a non-vanishing Berry curvature arising from the topologically non-trivial spin texture.[9-11] Such band structure also induced large anomalous Nernst effect,[12] topological Hall effect,[13] magneto-optical Kerr effect (MOKE)[7, 14, 15] and spin Hall effect[16] in $Mn_3Sn$ single crystals. Moreover, a non-zero Berry curvature over the occupied electronic states and a band exchange splitting can be observed even without net magnetization,[17, 18] making such materials particularly attractive for spintronic and magneto-optical device applications.[19-21]

Despite of the progress, most $Mn_3Sn$ materials studied to date are single crystals. Although a few groups reported successful epitaxial growth of $Mn_3Sn$ thin films on single crystal $Al_2O_3$ or MgO substrates by sputtering, the anomalous Hall effect and magneto-optical effect are inferior compared to single crystals.[13, 22] So far, it is still very challenging to fabricate high quality epitaxial $Mn_3Sn$ thin films. Meanwhile, even less reports are focused on the magneto-optical properties. It is unclear how does the magneto-optical property of $Mn_3Sn$ epitaxial thin films compare with their bulk counterparts. As an alternative method, PLD is capable of fabricating materials with complex components and non-equilibrium phases, which is ideal for exploring new topological magnetic materials. However, so far, there has been no reports on growth of Mn$_3$Sn epitaxial thin films by PLD to the best of our knowledge.

In this work, we report successful growth of high quality epitaxial Mn$_3$Sn thin films grown by PLD. The epitaxial Mn$_3$Sn films show large anomalous Hall effect up to 3.02 μΩ cm and strong LMOKE up to 38.1 mdeg at 633 nm wavelength at 300 K temperature, which are almost identical to Mn$_3$Sn single crystals. Our report shows that PLD is a promising method for growth of high-quality Mn$_3$Sn thin films, paving the way for future explorations on device applications based on WSMs.

## EXPERIMENTAL SECTION

### Fabrication of thin films

80 nm thick (0001)- and (11$\bar{2}$0)-oriented Mn$_{3+x}$Sn$_{1-x}$ films were deposited on single crystal Al$_2$O$_3$ (0001), (1$\bar{1}$02) and MgO (110) substrates by PLD. The background pressure was better than $1\times10^{-5}$ Pa. For Mn$_{3+x}$Sn$_{1-x}$ films growth on Al$_2$O$_3$ (0001) substrates, an epitaxial Pt buffer layer was first grown at 450 °C, and allowed to cool to room temperature. The (0001) oriented Mn$_{3+x}$Sn$_{1-x}$ films were then grown at room temperature on c-Al$_2$O$_3$ substrate on top of the Pt buffer layer, and in-situ annealed at 480 °C and holding for 1 hour. (11$\bar{2}$0) Mn$_{3+x}$Sn$_{1-x}$ films were grown at room temperature on R- Al$_2$O$_3$ and MgO (110) substrate without any seed layers, and in-situ annealed at 520 °C and 550 °C, holding for 1 hour. During the deposition, the fluence of the KrF excimer laser for Pt and Mn$_3$Sn growth was fixed at 3.0 J cm$^{-2}$ and 2.5 J cm$^{-2}$, respectively. The target-substrate distance of Pt and Mn$_3$Sn growth was fixed at 4.5 cm and 5.5 cm, respectively.

### Characterizations

The film compositions were measured by Rutherford backscattering spectrometry (RBS) and energy-dispersive spectroscopy (EDS) in a scanning electron microscope (SEM, JSM-7600F). The thickness of Mn$_{3+x}$Sn$_{1-x}$ film was measured by SEM. The crystal structure of the films was analyzed using X-ray diffraction (XRD) and reciprocal space mapping (RSM) measurements by a high-resolution X-ray diffractometer (Bruker

D8). The characterizations of surface morphology were performed using a commercial atomic force microscope (AFM) (Cypher Asylum Research). Magnetic hysteresis was measured from 2 K to 300 K temperature using a commercial superconducting quantum interference device (SQUID) magnetometer (MPMS, Quantum Design). Hall resistivity was measured by the Van-der-Pauw method from 2 K to 300 K temperature using a commercial cryogen free transport measurement system (mCFS, Cryogenic). The LMOKE hysteresis was measured at 300 K temperature using a free space MOKE measurement setup (MOKE system, East Changing Technologies). In the longitudinal MOKE configuration, the magnetic field was parallel to the plane of incidence and the reflecting surface. Linear polarized light beam was incident on the sample with an angle of 45°. The MOKE hysteresis was collected with in-plane applied magnetic field up to ±0.25 T.

## RESULTS AND DISCUSSION

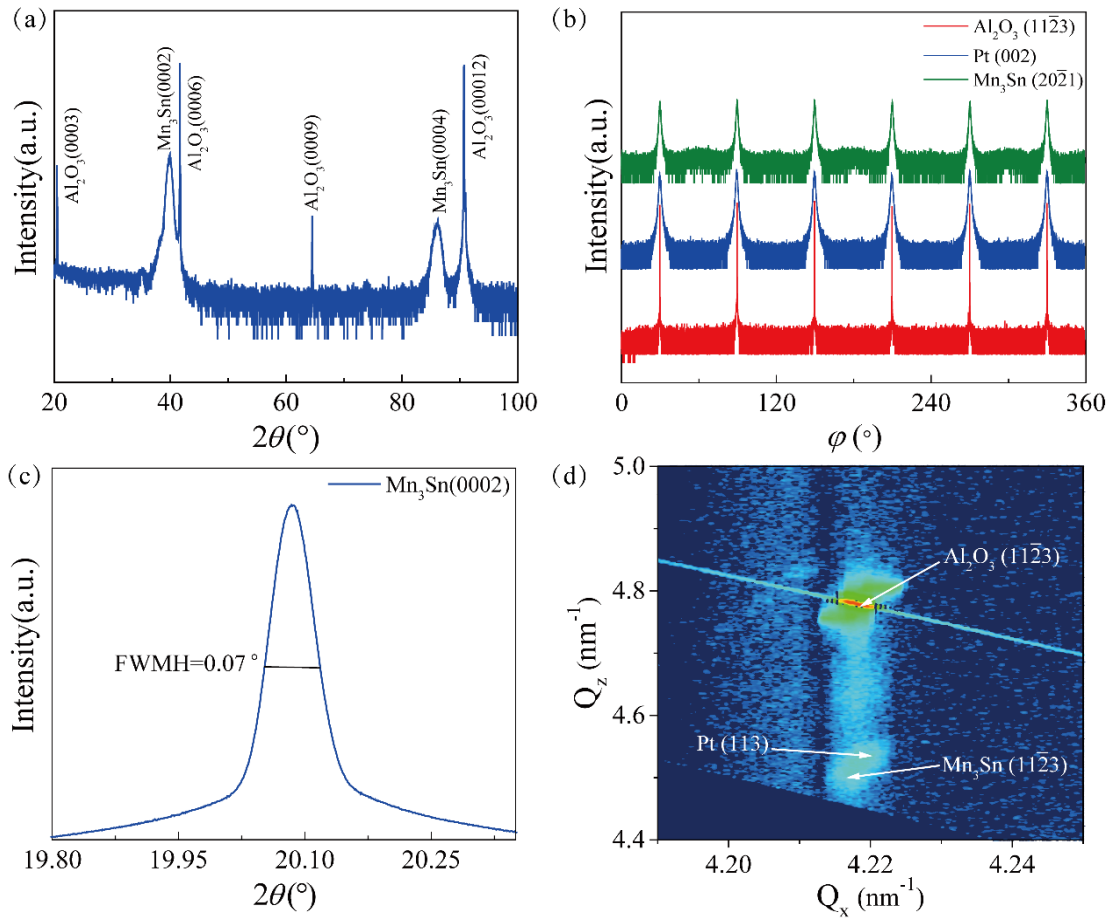

**Figure 1** Structure characterization of $Al_2O_3$/$Mn_3Sn$ (0001) films. (a) XRD $2\theta$-$\omega$ scan of the $Al_2O_3$/$Mn_3Sn$ (0001) film. (b) In-plane $\varphi$ scans of $Mn_3Sn$ (20$\bar{2}$1), Pt (002), and $Al_2O_3$ (11$\bar{2}$3) planes. (c) Rocking curve of the $Mn_3Sn$ (0002) peak. (d) Reciprocal-space mapping (RSM) of the $Mn_3Sn$ (11$\bar{2}$3), Pt (113) and $Al_2O_3$ (11$\bar{2}$3) planes, demonstrating coherent and epitaxial growth of $Mn_3Sn$ on $Al_2O_3$.

Figure 1a shows the $2\theta$-$\omega$ XRD pattern measured for $Mn_3Sn$ (0001) films deposited on $Al_2O_3$ (0001) substrates. The pattern shows strong $Mn_3Sn$ (0002) and (0004) diffraction peaks without any secondary phase diffraction. The Pt thin film diffraction peak is merged with the $Mn_3Sn$ peaks due to their similar lattice constants.[23,24] Figure 1b displays the in-plane $\varphi$ scans of the $Al_2O_3$ (11$\bar{2}$3), Pt (002), and $Mn_3Sn$ (20$\bar{2}$1) planes. Diffraction peaks with six fold symmetry confirm the epitaxial relation of $Al_2O_3$ [10$\bar{1}$0] ∥ Pt [110] ∥ $Mn_3Sn$ [2$\bar{1}\bar{1}$0]. The rocking curves show a very narrow peak with full width at a half maximum (FWHM) of 0.07°, comparable to the best epitaxial $Mn_3Sn$ films prepared by sputtering so far.[23] This result indicates that the film has exquisite planar alignment of the (0001) crystalline planes between the film and the substrate with a small d-spacing spread, as show in Figure 1c. Figure 1d illustrates the RSM of $Mn_3Sn$ (11$\bar{2}$3), Pt (113) and $Al_2O_3$ (11$\bar{2}$3) planes. A coherent interface is observed between the film and the substrate due to the same in-plane lattice constants between $Mn_3Sn$ and $Al_2O_3$. For a hexagonal crystal with in-plane and out-of-plane lattice constants $a$ and $c$, we can calculate the $a$ and $c$ lattice parameters of $Mn_3Sn$ to be 5.60 Å and 4.54 Å, respectively. This is in good agreement with $Mn_3Sn$ single crystals with $a$ = 5.67 Å and $c$ = 4.53 Å.[21] Therefore, the film is fully strained by in-plane biaxial compressive strain by forming the coherent interface with $Al_2O_3$.

Figure 2a shows the $2\theta$-$\omega$ XRD pattern of a $Mn_3Sn$ (11$\bar{2}$0) film grown on a R-$Al_2O_3$ substrate. The diffraction peaks of $Mn_3Sn$ (11$\bar{2}$0) planes can be clearly observed. Due to a large lattice mismatch of 4.79% between $Al_2O_3$ (1$\bar{1}$02) and $Mn_3Sn$ (11$\bar{2}$0) planes, it is very difficult to grow epitaxial $Mn_3Sn$ films with such a large lattice mismatch. We notice that accurate temperature control is very important for the growth of epitaxial Mn₃Sn films. For highly oriented Mn₃Sn film growth, we carefully controlled the temperature of in-situ annealing in PLD. As show in Figure S3, highly oriented films are only obtained in a narrow temperature window around 520 °C. In this film, we observe almost exclusive Mn₃Sn (11$\bar{2}$0) diffractions, with only a very weak peak at $2\theta = 40.7°$ originated from Mn₃Sn (0002) planes. A comparison between the $\varphi$ scan of the (20$\bar{2}$1) planes of the Mn₃Sn and the (11$\bar{2}$0) planes of Al₂O₃ is displayed in Figure 2b. The crystallographic orientation relation is determined as Al₂O₃ [11$\bar{2}$0] ∥ Mn₃Sn [2$\bar{1}\bar{1}$0]. An inspection of the $\varphi$ scan shows that the peaks are separated by 90° for the film with fourfold symmetry, which is different from the 180° separation of the $\varphi$ scan of the Al₂O₃ substrates. This result indicates two sets of Mn₃Sn domains with 90° rotation in-plane.[25] The RSM also indicates a semi-coherent interface between the film and the substrate. The lattice constants of this Mn₃Sn film are $a$= 5.66 Å and $c$=4.58 Å and, respectively.

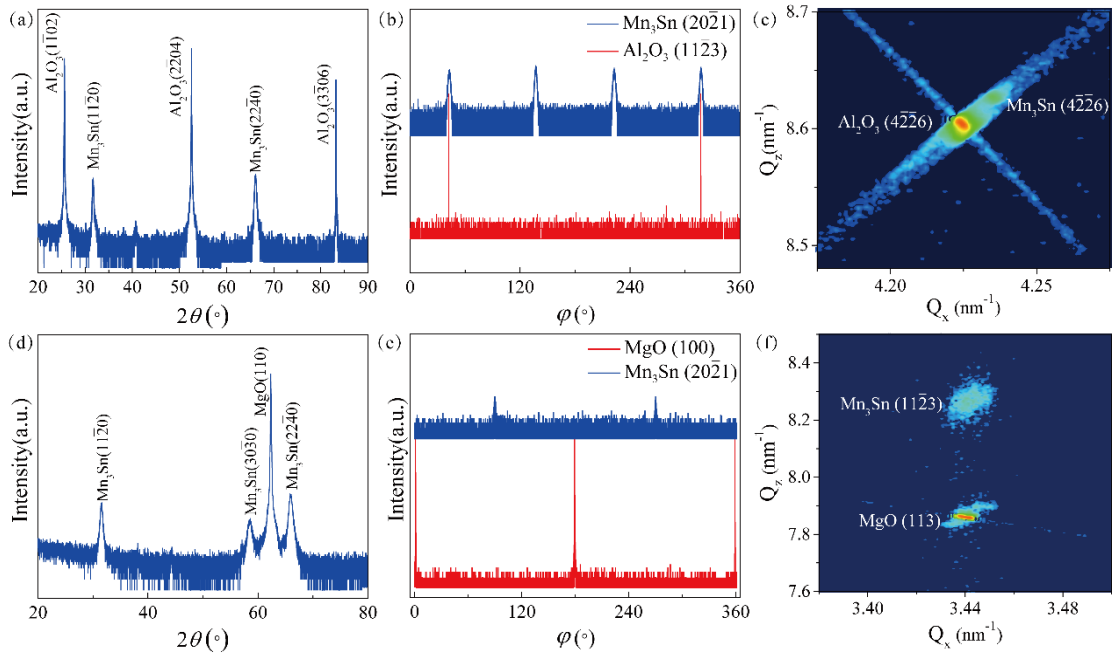

**Figure 2** Structure characterization of Al₂O₃/Mn₃Sn (11$\bar{2}$0) and MgO/Mn₃Sn (11$\bar{2}$0) films. (a) and (d) XRD $2\theta$-$\omega$ scan of the Al₂O₃/Mn₃Sn (11$\bar{2}$0) and MgO/Mn₃Sn (11$\bar{2}$0) films, respectively. (b) In-plane φ scans of the Mn₃Sn (20$\bar{2}$1), and Al₂O₃ (11$\bar{2}$3) planes. (c) RSM of the Mn₃Sn (4$\bar{2}\bar{2}$6) and Al₂O₃ (4$\bar{2}\bar{2}$6) planes, demonstrating semi-coherent

and epitaxial growth. (e) In-plane $\varphi$ scans of Mn$_3$Sn (20$\bar{2}$1), and MgO (100) planes. (f) RSM of the Mn$_3$Sn (11$\bar{2}$3) and MgO (113) planes, demonstrating coherent and epitaxial growth.

Figure 2d shows the $2\theta$-$\omega$ XRD pattern of the Mn$_3$Sn (11$\bar{2}$0) films grown on MgO (110) substrates. Despite of a large lattice mismatch of 4.9% between MgO (110) and Mn$_3$Sn (11$\bar{2}$0) films, epitaxial growth is still achieved, except for only a small peak from the Mn$_3$Sn (30$\bar{3}$1) diffraction plane. A comparison of the $\varphi$ scans obtained from the (20$\bar{2}$1) planes of the Mn$_3$Sn films and from (100) planes of MgO substrates are displayed in Figure 2e. Four diffraction peaks from Mn$_3$Sn and two from MgO are observed, confirming the epitaxial relation of MgO [110] ∥ Mn$_3$Sn [2$\bar{1}\bar{1}$0]. The RSM shown in Figure 2f also demonstrates the lattice constants for Mn$_3$Sn are $a$=5.67 Å and $c$=4.53 Å, respectively. In this case, the Mn$_3$Sn film is fully strained in-plane by the MgO substrate. The Mn$_3$Sn epitaxial films grown on different substrates show different lattice constants, facilitating strain and magnetic anisotropy engineering of this material for future device applications.[22]

Next, we characterized the magnetic hysteresis loops at different temperatures of the Mn$_3$Sn films by SQUID magnetometry, as shown in Figure 3. In this figure, the diamagnetic signals from the substrates were first measured and subtracted from the samples (see un-subtracted M-H curves in Figure S1). Weak ferromagnetic hysteresis is observed at $T$ = 300 K, indicating an antiferromagnetic phase of the deposited thin films. The ferromagnetic signal is originated from the canted magnetic moments in the a-b plane, showing uncompensated moment of the noncollinear antiferromagnetic structure.[6, 26] The saturation magnetization of Mn$_3$Sn increases with decreasing temperature as shown in Figure 3d, indicating a phase transition from the chiral antiferromagnetic phase to the glassy ferromagnetic phase.[26, 27] The magnetic properties of Mn$_3$Sn are strongly related to the crystal orientations, showing anisotropic hysteresis of the magnetization curve $M$ ($B$) for different samples. At 300 K temperature, the 80 nm Al$_2$O$_3$/Mn$_3$Sn (0001), Al$_2$O$_3$/Mn$_3$Sn (11$\bar{2}$0) and MgO/Mn$_3$Sn (11$\bar{2}$0) films show out-of-plane saturation magnetizations of $M = 1.85$ emu/cm$^3$, $M = 14.63$ emu/cm$^3$ and $M = 21.5$ emu/cm$^3$, and coercivities of $H_c$=276 Oe, 20.2 Oe and 243.5 Oe, respectively. This result is comparably higher than previous reports of sputtered Mn$_3$Sn epitaxial films.[24] The larger magnetization observed in MgO/Mn$_3$Sn (11$\bar{2}$0) compared to Al$_2$O$_3$/Mn$_3$Sn (11$\bar{2}$0) film can be attributed to additional uncompensated magnetic moments arising from excess Mn atoms. The stoichiometry of these films is measured by RBS and EDS, showing a chemical formula of Mn$_{3.2}$Sn$_{0.8}$ for Al$_2$O$_3$/Mn$_3$Sn (11$\bar{2}$0) and Mn$_{3.4}$Sn$_{0.6}$ for MgO/Mn$_3$Sn (11$\bar{2}$0) films respectively. The excess Mn could randomly occupy the Sn sites, introducing excess magnetic moments.[23] As decreasing temperature, both the saturation magnetization and coercivity increase in all three samples. As shown in Fig. 3d, the M$_s$ of the Al$_2$O$_3$/Mn$_3$Sn (11$\bar{2}$0) film reaches 53.9 emu/cm$^3$ at 2 K, matching with the single crystals.[31] Whereas the M$_s$ of MgO/Mn$_3$Sn (11$\bar{2}$0) film shows low M$_s$ of 35.4 emu/cm$^3$ above 200 K, and high M$_s$ up to 87.6 emu/cm$^3$ at 50 K. (we excluded the lower temperature data due to a clear paramagnetic signal from the MgO substrate)

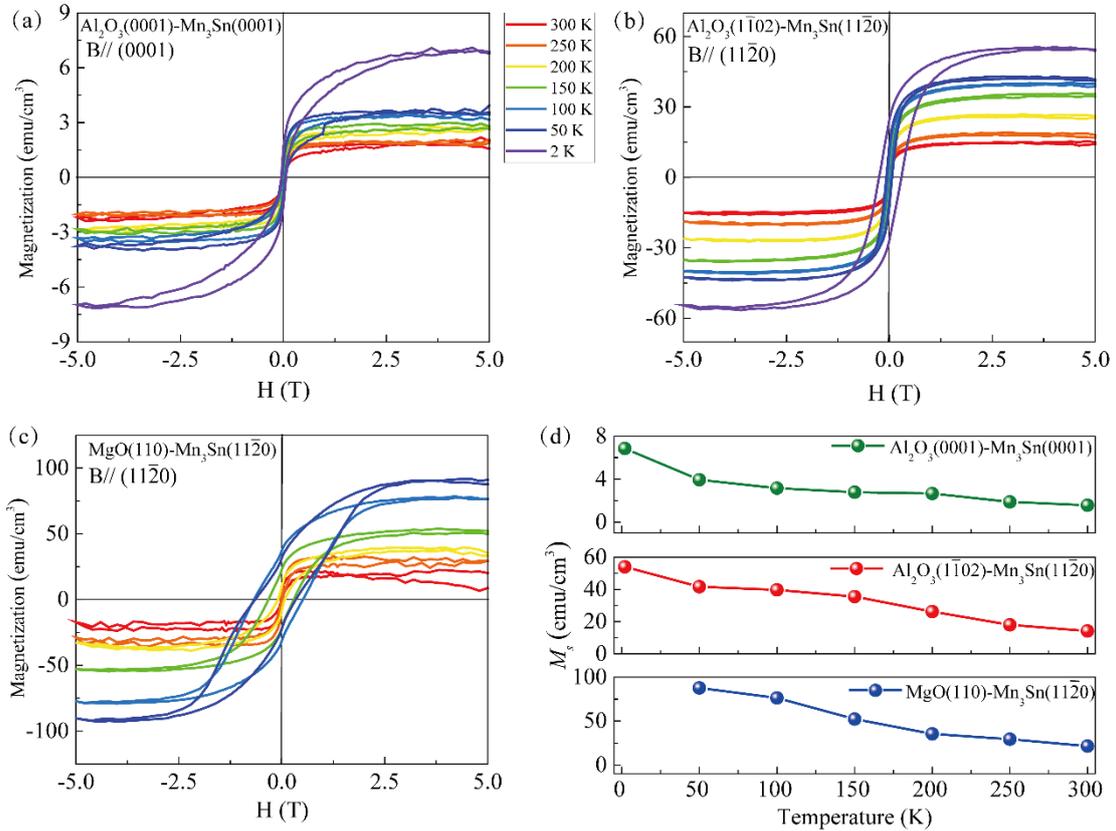

**Figure 3** Temperature dependent magnetic hysteresis of Mn$_3$Sn films (a) M-H curves measured between 2–300 K for the Al$_2$O$_3$/Mn$_3$Sn (0001) film. (b) M-H curves measured at 2–300 K for the Al$_2$O$_3$/Mn$_3$Sn (11$\bar{2}$0) film. (c) M-H curves measured at 50–300 K for the MgO/Mn$_3$Sn (11$\bar{2}$0) film. (d) M$_s$-T curves as a function of temperature for the three Mn$_3$Sn films.

Next, we measured the Hall effect on Mn$_3$Sn films at different temperatures, as shown in Figure 4. The magnetic field is applied along the surface normal direction for all films, corresponding to the [0001] (magnetic difficult axis) or [11$\bar{2}$0] (magnetic easy axis) directions for different Mn$_3$Sn films. The Hall resistance R$_{xy}$ shows a nonlinear feature as a function of applied magnetic field, as shown in Figure 4a. A hysteresis loop of Al$_2$O$_3$/Mn$_3$Sn (0001) film with a sizable jump of $|\Delta\rho_H| \approx 0.053 \, \mu\Omega$ cm is observed at 300 K. By contrast, we can see a hysteresis loop from the Al$_2$O$_3$/Mn$_3$Sn (11$\bar{2}$0) film with a much larger jump of $|\Delta\rho_H| \approx 3.07 \, \mu\Omega$ cm at 300 K, as shown in Figure 4b. This result is comparably higher than previous reports of sputtered Mn$_3$Sn epitaxial films.[23, 24, 27] It is also approximately the same compared to bulk Mn$_3$Sn single crystals, indicating a high quality of our Mn$_3$Sn films fabricated by PLD.[6] The saturated Hall resistivity at positive applied magnetic field is negative, indicating intrinsic contribution of the Berry curvature in *k*-space.[2, 29] For Mn$_3$Sn on Al$_2$O$_3$ films shown in Figure 4a and 4b, as temperature decreases, $|\Delta\rho_H|$ first increases, then decreases when the sample temperature is below 250 K. This phenomenon indicates a phase transition from the Kagome lattice at high temperatures to spin spiral/glass at low temperatures. The thin film phase transition temperature of around 250 K matches well with the phase transition temperature of 225 K in bulk Mn$_3$Sn single crystals.[6,13] Different temperature dependence of the Hall effect is observed in MgO/Mn$_3$Sn (11$\bar{2}$0) films. As shown in Figure 4c, a hysteresis loop of MgO/Mn$_3$Sn (11$\bar{2}$0) film with a jump of $|\Delta\rho_H| \approx 0.2 \, \mu\Omega$ cm is observed at 300 K. With decreasing the temperature, the $\rho_H$ reaches around $0.4 - 0.7 \, \mu\Omega$ cm, lower than that observed in the Al$_2$O$_3$/Mn$_3$Sn (11$\bar{2}$0) sample. The weaker AHE may be caused by the higher Mn concentration in MgO/Mn$_3$Sn (11$\bar{2}$0) of Mn$_{3.4}$Sn$_{0.6}$. The higher Mn concentration may lead to a gap opening effect, which tends to annihilate the Weyl nodes and greatly reduce the Berry curvature.[23] The $\rho_H$ is negative for the whole temperature range. A maximum $\rho_H$ is observed at 150 K. One possibility of the lower phase transition temperature may be attributed to intrinsic magnetic reversal behavior and existence of the heterogeneous microstructure.[30, 31] Another possibility is that the Sn sites are substituted by extra Mn atoms, leading to a lower phase transition temperature, making a significant impact on electrical performance at low temperature.[32] Figure 4d plots the AHE as a function of temperature for all three samples. The temperature dependence of $\rho_H$ for Al$_2$O$_3$/Mn$_3$Sn(11$\bar{2}$0) in the middle panel closely resembles the report on single crystals.[23] However, this curve for MgO/Mn$_3$Sn (11$\bar{2}$0) shows a different trend with the largest $\rho_H$ observed at 150 K, and a finite $\rho_H$ of -0.4 μΩ•cm at 2 K. The transition temperature of ~150 K also matches with the magnetic hysteresis in Figure 3c. Such behavior was not observed in previous samples. Therefore, the band structure and spin structure of this film needs further investigation in the future.

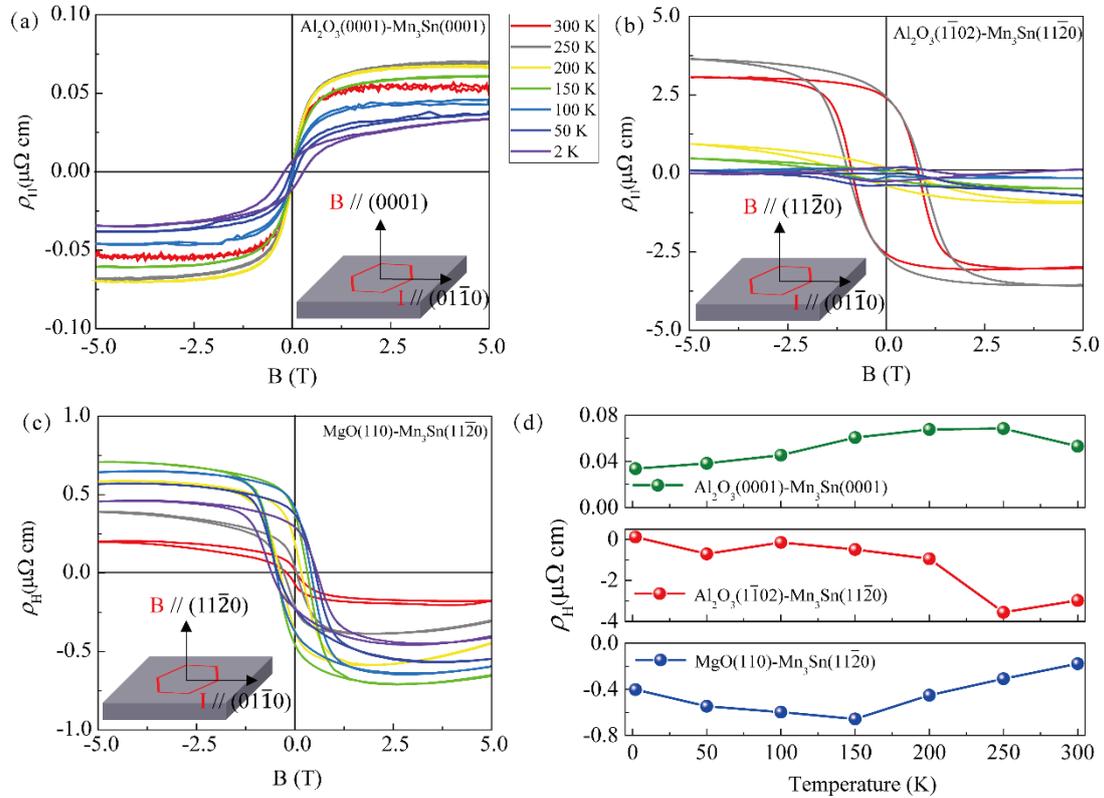

**Figure 4** Hall resistivity measured as a function of out of plane applied magnetic field

at different temperatures for (a) Al$_2$O$_3$/Mn$_3$Sn (0001) (b) Al$_2$O$_3$/Mn$_3$Sn (11$\bar{2}$0) and (c) MgO/Mn$_3$Sn (11$\bar{2}$0) films. (d) Anomalous Hall resistivity as a function of temperature for different Mn$_3$Sn films.

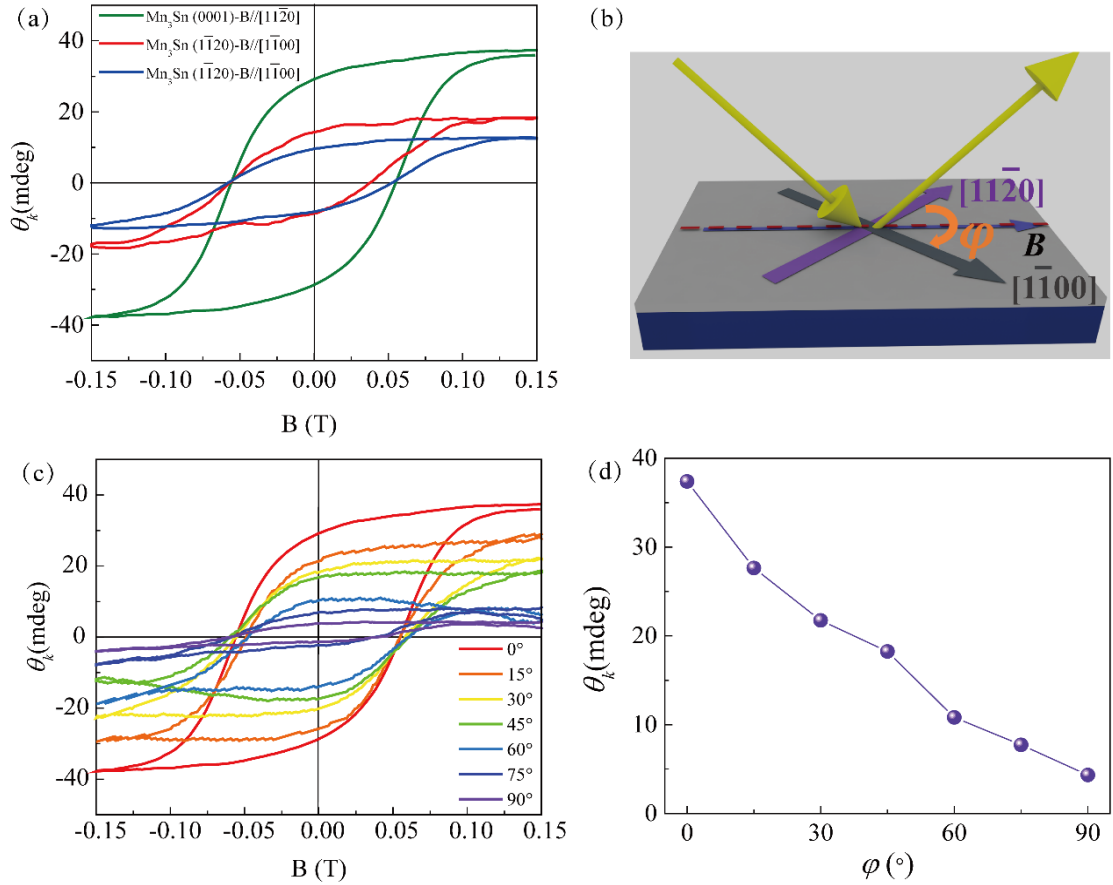

**Figure 5** Longitudinal magneto-optical Kerr hysteresis of the Mn$_3$Sn film measured at 300 K. (a) LMOKE for Al$_2$O$_3$/Mn$_3$Sn (0001), Al$_2$O$_3$/Mn$_3$Sn (11$\bar{2}$0) and MgO/Mn$_3$Sn (11$\bar{2}$0) films. (b) Schematic illustrations of sample configurations for in-plane anisotropic LMOKE characterization. (c) The anisotropic LMOKE hysteresis in the (0001) plane of the Al$_2$O$_3$/Mn3Sn (0001) sample. (d) LMOKE saturation rotation angle $\theta_K$ as a function of the in-plane azimuth angle $\varphi$.

To study the magneto-optical effect of the Mn$_3$Sn films, LMOKE hysteresis was measured at 633 nm wavelength at 300 K temperature. Large anisotropic MOKE was observed in previous reports on Mn$_3$Sn single crystals due to the octupole polarization induced non-zero Berry curvature.[7] But the reports on thin films were rare.[14] Figure 5a shows the LMOKE hysteresis measured under 45° incident angle. The applied in-plane magnetic fields are parallel to the [11$\bar{2}$0] crystal orientation of the Al$_2$O$_3$/Mn$_3$Sn (0001) sample, and to the [1$\bar{1}$00] crystal orientation of the Al$_2$O$_3$/Mn$_3$Sn (11$\bar{2}$0) and MgO/Mn$_3$Sn (11$\bar{2}$0) samples respectively. For all samples, clear LMOKE square hysteresis loops are observed. Large saturation Kerr rotation angles of $|\theta_K| = 37.4$ mdeg, $|\theta_K| = 18.3$ mdeg and $|\theta_K| = 12.8$ mdeg are observed for the Al$_2$O$_3$/Mn$_3$Sn (0001), Al$_2$O$_3$/Mn$_3$Sn (11$\bar{2}$0) and MgO/Mn$_3$Sn (11$\bar{2}$0) films, respectively. These values matches with the highest LMOKE values reported so far on Mn$_3$Sn single crystals, further proving the high quality of our films.[33] We also characterized the dependence of the LMOKE hysteresis on the in-plane azimuth angle $\varphi$ as sketched in Figure 5b. By rotating the directions of the applied magnetic field $B$ from [11$\bar{2}$0] ($\varphi = 0°$) to [1$\bar{1}$00] ($\varphi = 90°$) in the (0001) plane, the LMOKE signal decreases from 38.1 mdeg to 4.3 mdeg, as shown in Figure 5c, d. The crystalline anisotropy of LMOKE also matches with the observation on Mn$_3$Sn single crystals.[7]

**CONCLUSIONS**

In summary, we have successfully grown epitaxial thin films of Mn$_3$Sn with (0001) and (11$\bar{2}$0) orientation on Al$_2$O$_3$ and MgO single crystal substrates by PLD. Highly oriented epitaxial Mn$_3$Sn films were demonstrated by X-ray diffraction. The temperature dependence of the magnetization and AHE of the Mn$_3$Sn films demonstrates their spin structure changing from chiral antiferromagnetic to spin/spiral glass structures during cooling, with the M-T and $\rho_H$-T curves strongly dependent on the film stoichiometry and substrate materials. Berry curvature induced large and anisotropic AHE and LMOKE are observed in the WSM phase of Mn$_3$Sn at 300 K temperature, which quantitatively match with the measurement results on bulk Mn$_3$Sn single crystals. Our results demonstrate high quality magnetic WSM epitaxial thin films can be fabricated by pulsed laser deposition, paving the way for spintronic and photonic device applications.

## ASSOCIATED CONTENT

**Supporting Information.**

The following files are available free of charge on the ACS Publications website at (to be inserted by the publisher).

The magnetization hysteresis loops with background of substrate; In-situ annealing dependence of XRD pattern of $Al_2O_3(1\bar{1}02)$ / $Mn_3Sn$ $(11\bar{2}0)$ films; RSM of the bare substrates.

## AUTHOR INFORMATION

**Corresponding Author**

**Ningbin Zhang** – Department of Materials Science and Engineering, City University of Hong Kong, Tat Chee Avenue 83, Kowloon, Hong Kong 999077, China; Shenzhen Futian Research Institute, City University of Hong Kong, Shenzhen 518048, China;

*Email: nbzhang13s@alum.imr.ac.cn

**Lei Bi** – National Engineering Research Center of Electromagnetic Radiation Control Materials, University of Electronic Science and Technology of China, Chengdu 610054, China; School of Electronic Science and Engineering, University of Electronic Science and Technology of China, Chengdu 610054, China;

*Email: bilei@uestc.edu.cn.

**Authors**

**Dong Gao** – National Engineering Research Center of Electromagnetic Radiation Control Materials, University of Electronic Science and Technology of China, Chengdu 610054, China; School of Electronic Science and Engineering, University of Electronic Science and Technology of China, Chengdu 610054, China;

**Zheng Peng** – National Engineering Research Center of Electromagnetic

Radiation Control Materials, University of Electronic Science and Technology of China, Chengdu 610054, China; School of Electronic Science and Engineering, University of Electronic Science and Technology of China, Chengdu 610054, China;

**Yunfei Xie** – National Engineering Research Center of Electromagnetic Radiation Control Materials, University of Electronic Science and Technology of China, Chengdu 610054, China; School of Electronic Science and Engineering, University of Electronic Science and Technology of China, Chengdu 610054, China;

**Yucong Yang** – National Engineering Research Center of Electromagnetic Radiation Control Materials, University of Electronic Science and Technology of China, Chengdu 610054, China; School of Electronic Science and Engineering, University of Electronic Science and Technology of China, Chengdu 610054, China;

**Weihao Yang** – National Engineering Research Center of Electromagnetic Radiation Control Materials, University of Electronic Science and Technology of China, Chengdu 610054, China; School of Electronic Science and Engineering, University of Electronic Science and Technology of China, Chengdu 610054, China;

**Shuang Xia** – National Engineering Research Center of Electromagnetic Radiation Control Materials, University of Electronic Science and Technology of China, Chengdu 610054, China; School of Electronic Science and Engineering, University of Electronic Science and Technology of China, Chengdu 610054, China;

**Wei Yan** – National Engineering Research Center of Electromagnetic Radiation Control Materials, University of Electronic Science and Technology of China, Chengdu 610054, China; School of Electronic Science and Engineering, University of Electronic Science and Technology of China, Chengdu 610054, China;

**Longjiang Deng** – National Engineering Research Center of Electromagnetic Radiation Control Materials, University of Electronic Science and Technology of China, Chengdu 610054, China;

**Tao Liu** – National Engineering Research Center of Electromagnetic Radiation Control Materials, University of Electronic Science and Technology of China, Chengdu 610054, China; School of Electronic Science and Engineering, University of Electronic Science and Technology of China, Chengdu 610054, China;

**Jun Qin** – National Engineering Research Center of Electromagnetic Radiation Control Materials, University of Electronic Science and Technology of China, Chengdu 610054, China; School of Electronic Science and Engineering, University of Electronic Science and Technology of China, Chengdu 610054, China;

**Xiaoyan Zhong** – Department of Materials Science and Engineering, City University of Hong Kong, Tat Chee Avenue 83, Kowloon, Hong Kong 999077, China; Shenzhen Futian Research Institute, City University of Hong Kong, Shenzhen 518048, China;

**Notes**

The authors declare no competing financial interest.

ACKNOWLEDGMENT

The authors are grateful for support by the Ministry of Science and Technology of the People's Republic of China (MOST) (Grant Nos. 2018YFE0109200 and 2021YFB2801600), National Natural Science Foundation of China (NSFC) (Grant No.s 51972044, 52021001, 52171014 and 52011530124), Sichuan Provincial Science and Technology Department (Grant No. 2019YFH0154 and 2021YFSY0016) and the Fundamental Research Funds for the Central Universities (Grant No. ZYGX2020J005). Science, Technology and Innovation Commission of Shenzhen Municipality (HZQB-KCZYB-2020031, SGDX20210823104200001, JCYJ20210324134402007), Sino-German Center for Research Promotion (M-0265), RGC of Hong Kong SAR, China (E-CityU101/20, G-CityU102/20, 11302121, 11309822), European Research Council (856538, "3D MAGiC"), CityU SIRG (7020016, 7020043), City University of Hong Kong (Projects no. 9610484, 9680291, 9678288) and the City University of Hong Kong Shenzhen Research Institute.